\documentstyle[epsfig,amsbsy,amssymb]{elsart}

\begin{document}
\begin{frontmatter}

\rightline{nucl-th/9712070}

\title{\bf Asymmetry in $\vec{\omega}$ meson photoproduction  \\
       and the phase of $\vec{\omega\pi\gamma}$ coupling}

\author[JINR]{Alexander I. Titov\thanksref{titov}},
\author[TUM]{Yongseok Oh\thanksref{yoh}}

\address[JINR]{Bogoliubov Laboratory of Theoretical Physics, JINR,
               141980 Dubna, Russia}
\address[TUM]{Institut f\"ur Theoretische Physik, Physik Department,
              Technische Universit\"at M\"unchen, D-85747 Garching,
              Germany}

\thanks[titov]{E-mail address : {\tt atitov@thsun1.jinr.ru}}
\thanks[yoh]{Alexander von Humboldt Fellow.
             Address after Oct. 1, 1997:
             Research Institute for Basic Sciences and
             Department of Physics, Seoul National University,
             Seoul 151-742, Korea.
             E-mail address : {\tt yoh@phya.snu.ac.kr}}

\begin{abstract}

We analyze the double polarization asymmetry of the $\omega$-meson
photoproduction in the vector-meson--dominance model of diffractive
production and the one-pion exchange model.
We find that the longitudinal beam-target asymmetry is very sensitive
to the real part of the diffractive photoproduction amplitude and to
the sign of the $\omega\pi\gamma$ coupling because of the different
spin structures of the amplitudes associated with the different
mechanisms.

\end{abstract}


\end{frontmatter}

\newpage

Most phenomenological models to the pion photoproduction include hadron
resonances as explicit degrees of freedom in addition to the low energy
theorem contributions \cite{OO,GG93,FM97} in order to explain the
experimental data including resonance region.
This inevitably leads us to the inclusion of the vector-meson degrees of
freedom as well as the nucleon resonances.
In some literature \cite{GG93,FM97,ITC}, it has been claimed that some
physical quantities of the $\pi^0$ photoproduction depend on the phases
of $V\pi\gamma$ coupling constants, where $V$ stands for a vector-meson,
and that it requires a special choice on the phases of $g^{}_{V\pi\gamma}$'s
to reproduce the experimental observation.
This problem has been also raised from the meson exchange
current contribution in the deuteron form factors \cite{GHSMS} and
proton-nucleon bremsstrahlung \cite{EGJF}.
Actually, in this case one deals with the relative phase of the 
vector-meson exchanged amplitudes and the Born terms which is chosen
often by the Gari and Hyuga convention \cite{GHSMS}.
However, these contributions are combined with other
mechanisms and the present level of experimental accuracy does not allow
us to fix the signs uniquely and the conclusion is still controversial.
Thus an independent process that is directly proportional to the phase
of the couplings will be useful to resolve this problem.

Another motivation of our study concerns with the real part of the
vector-meson--dominance model (VDM) amplitude, which is related to
the study of the strangeness content of the nucleon through the
$\phi$-meson photoproduction from proton \cite{TYO97}.
In Ref. \cite{TOY97} it is pointed out that the double polarization
observables in $\phi$ photoproduction may be used as a good probe for
hidden strangeness in the proton.
Although its effect is expected to be small, however, the interference
between the real part of the VDM amplitude and the one-pion exchange
model (OPE) amplitude may give some contribution to the asymmetries.
Therefore, independent analyses on the real part of the VDM amplitude
are highly desirable.

In this paper we propose to use the double polarization asymmetry
of $\omega$ photoproduction as a tool to resolve the above two issues.
We will show that the beam--target double polarization asymmetry of
$\omega$ photoproduction is proportional to the product of
$g_{\omega\pi\gamma}$ and the real part of VDM amplitude and that the
measurements of the asymmetry can give an information about the two
quantities directly, because the real part of the VDM amplitude is
strictly related to it's imaginary part by the dispersion relation 
while the phase of the imaginary part is fixed by the unitarity condition.
Throughout this paper, we restrict our consideration to the
$\omega\pi\gamma$ coupling because the asymmetries in the $\rho$
photoproduction are expected much smaller because of the relatively
small value of the $\rho\pi\gamma$ coupling constant.

\begin{figure}[t]
\centering
\epsfig{file=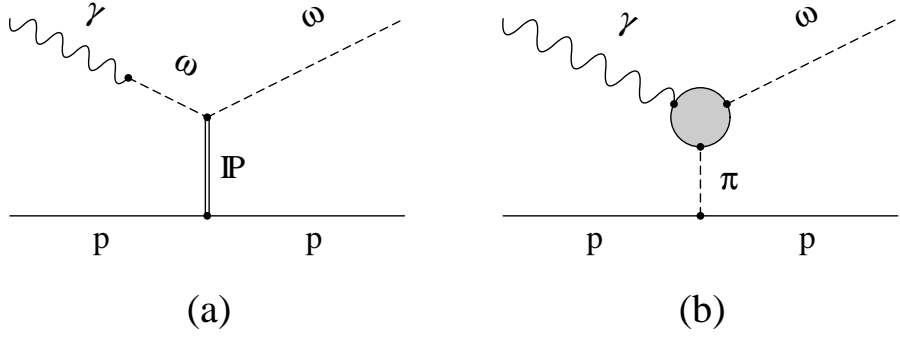, width=12cm}
\caption{(a) Diffractive $\omega$-meson photoproduction within the
         vector-meson-dominance model by means of a Pomeron
         ($\mathbb{P}$) exchange.
         (b) One pion exchange process in the $\omega$ photoproduction.
         The blob includes the direct $\omega\pi\gamma$ coupling and the
         $\omega\pi\rho$ coupling with $\rho\gamma$ vertex.
}
\label{fig:prod}
\end{figure}

Following Ref. \cite{Ballam73}, we assume that, for the initial photon
energy in a few GeV region, the total amplitude of vector-meson
photoproduction process comes mainly from two sources:
the vector-meson dominance model of diffractive production and the
one-pion exchange model as depicted in Fig. \ref{fig:prod}.
We define the kinematical variables as follows.
The four-momenta of the incoming photon, outgoing $\omega$, initial
proton and final proton are $k$, $q$, $p$ and $p'$, respectively.
In laboratory frame, we define
$k = (E_\gamma^L, \vec{k}_L)$, $q = (E_\omega^L, \vec{q}_L)$,
$p = (E_p^L, \vec{p}_L)$ and $p' = (E_{p'}^L, \vec{p}_L')$.
The corresponding variables in CM system are defined as
$k = (\nu, \vec{k})$, $q = (E_\omega, \vec{q})$,
$p = (E_p, -\vec{k})$ and $p' = (E_{p'}, -\vec{q})$.
We also use $t = (p - p')^2$ and $s = (p + k)^2$ with the nucleon mass
$M_N$, the $\omega$-meson mass $M_\omega$ and the pion mass $M_\pi$.

The diffractive $\omega$ photoproduction mechanism of VDM assumes that
the incoming photon mixes into the $\omega$-meson and then scatters
diffractively with the proton through an exchange of a Pomeron
($\mathbb{P}$) \cite{VDM}.
Experimental observations for the vector-meson production, small-$|t|$
elastic scattering and diffractive dissociation indicate that the
Pomeron behaves rather like an isoscalar photon-like particle \cite{DL1}.
Although it is widely accepted that the Pomeron can be described in terms
of non-perturbative two gluon exchange \cite{DL,Cudel,Gol93,Laget}, in
this paper we make use of the Pomeron--photon analogy, which is expected
to be valid at low energy.
Using the spin structure of the standard $VV\gamma$ coupling \cite{BD}
for the $VV\mathbb{P}$ vertex, the invariant amplitude of the diffractive
production process reads
\begin{eqnarray}
T_{m_f,\lambda_\omega;m_i,\lambda_\gamma}^{\rm VDM} &=&
 i\, T_{0}\, {\bar u}(p',m_f) \gamma_\alpha u(p,m_i) \,
 \varepsilon_\mu^{\lambda_\omega*}(\omega) 
\Gamma^{\alpha,\mu\nu} \varepsilon_\nu^{\lambda_\gamma}(\gamma),
\nonumber \\
\Gamma^{\alpha,\mu\nu} &=&
(k+q)^\alpha\, g^{\mu\nu} - k^{\mu}\, g^{\alpha\nu}
  - q^{\nu}\, g^{\alpha\mu},
\end{eqnarray}
where $m_{i(f)}$ is the spin projection of the initial (final) proton and
$\lambda_{\gamma(\omega)}$ is the helicity of the photon ($\omega$-meson).
Here $T_0$ includes the dynamics of Pomeron-hadron interaction,
$\varepsilon_\mu(\omega)$ and $\varepsilon_\mu(\gamma)$ are the
polarization vectors of the $\omega$ and the photon, respectively,
and $u(p)$ is the proton Dirac spinor.
We use the form and parameters of $T_0$ determined from the
parameterization \cite{Ballam73},
\begin{eqnarray}
\left( \frac{d\sigma}{dt} \right)_{\rm VDM} =
  c \left( 1+\frac{d}{E_\gamma} \right) \exp( b_\omega \, t),
\label{dsdt:vdm}
\end{eqnarray}
with $b_\omega = 6.7$ GeV$^{-2}$, $c=9.3$ $\mu$b/GeV$^2$
and $d=1.4$ GeV which are determined from the experimental data at
$\sqrt{s} = 2.8 \sim 9.7$ GeV as in Ref. \cite{Ballam73}.
The phase of $T^{\rm VDM}$ is assumed to be fixed by the optical theorem.

The relevant amplitude of the OPE diagram reads \cite{JMLS}
\begin{eqnarray}
T_{m_f,\lambda_\omega;m_i,\lambda_\gamma}^{\rm OPE} =
 \frac{1}{t - M_\pi^2} g_{NN\pi}
\tilde g_{\omega\pi\gamma} W^F_{m_f,m_i}
W^B_{\lambda_\omega,\lambda_\gamma},
\end{eqnarray}
where
\begin{eqnarray}
W^F_{m_f,m_i} = \bar u(p',m_f) \gamma_5 u(p,m_i),  \qquad
W^B_{\lambda_\omega,\lambda_\gamma} =
        i \epsilon^{\mu\nu\alpha\beta} q_\mu k_\alpha
        \varepsilon_\nu^{\lambda_\omega *} (\omega) 
	\varepsilon_\beta^{\lambda_\gamma} (\gamma),
\end{eqnarray}
from the interaction Lagrangian,
\begin{equation}
{\cal L}_{\omega\gamma\pi} = \tilde g_{\omega\pi\gamma}
\epsilon^{\mu\nu\alpha\beta}
\partial_\mu \omega_\nu \partial_\alpha A_\beta \pi^0,
\end{equation}
with the photon field $A_\beta$.
Direct calculation of $W^F$ and $W^B$ gives
\begin{eqnarray}
W^F_{m_f,m_i} &=& C \left\{
2 m_f \, (\alpha' \cos\theta - \alpha) \, \delta_{m_f,m_i}
- \alpha' \sin\theta \, \delta_{m_f,-m_i}
\right\},  \\
W^B_{\lambda_\omega,\lambda_\gamma} &=&
- \nu \left\{
\lambda_\gamma (E_\omega- q\cos\theta)\,\,
               \vec{\varepsilon}(\omega) \cdot \vec{\varepsilon}(\gamma)
+ \frac{q\sin\theta}{\sqrt{2}M_\omega}
(q-E_\omega\cos\theta) \, \delta_{\lambda_\omega,0}
\right. \nonumber \\ && \qquad \left. \mbox{}
-\frac{1}{2} \lambda_\omega q\sin^2\theta \right\},
\end{eqnarray}
where
\begin{eqnarray}
\vec{\varepsilon} (\omega) \cdot \vec{\varepsilon} (\gamma) =
 \left[ 1+ \left( \frac{E_\omega}{M_\omega} -1 \right)
        \delta_{\lambda_\omega,0} \right]
  d^1_{\lambda_\gamma,\lambda\omega}(\theta),
\end{eqnarray}
with $\theta$ the scattering angle in CM frame, $q \equiv |\vec{q}|$ and
$C=\sqrt{(\gamma_p+1)(\gamma_p'+1)}/2$.
We use $\gamma_p = E_p/M_N$ and $\alpha=\sqrt{(\gamma_p - 1)/(\gamma_p + 1)}$
for the initial proton while $\gamma_p'$ and $\alpha'$ are defined in the
same way for the final proton.
Note that the OPE amplitude is {\it pure real\/}.
We use $g^2_{NN\pi}/4\pi = 12.562$ and the effective coupling constant
$\tilde g_{\omega\pi\gamma}$ is determined
from the decay of $\omega \to (\rho \pi) \to \gamma \pi$ so that
${\tilde g}_{\omega\pi\gamma}^2 = 0.498$ GeV$^{-2}$.
Each vertex contains the Benecke-D\"urr form factors as given in
Refs. \cite{JMLS,BD68}.

In the literature, the Pomeron exchange amplitude is mostly assumed to
be pure imaginary \cite{Pomeron}.
In this approximation, OPE amplitude does not interfere with the VDM one
in the cross section.
However, if we assume the real part of the VDM amplitude, then it may
interfere with the OPE amplitude.
The real part of the VDM amplitude may be estimated using the subtracted
dispersion relation for the amplitude $f(s,t)$ normalized so that
$s \sigma_T = \mbox{Im}\, f(s,|t|_{\rm min})$ as in Ref. \cite{Bronzan},
\begin{eqnarray}
\mbox{Re} f(s,t) = \frac{2s^2}{\pi} \,\,\mbox{P} \int_{s_{\rm min}}^\infty
\frac{{\d} s'}{s'({s'}^2-s^2)} \mbox{Im}\, f(s',t),
\label{dr}
\end{eqnarray}
which can be evaluated analytically to give a derivative relation in the
limit of high energy.
However, at finite energy region of our interest, we need to evaluate
(\ref{dr}) in a numerical way.
Since the parameterization of (\ref{dsdt:vdm}) is valid only in the
limited region of energy and we have to integrate (\ref{dr}) over the
whole region of $s$, we assume the standard $s$-dependence of the
imaginary part as $f \sim s^{\alpha^{}_P}$ with $\alpha^{}_P \simeq 1$
instead of using Eq. (\ref{dsdt:vdm}).
This gives us the ratio
${\cal R} \equiv \mbox{Re}\, f(s,t) / \,\mbox{Im}\, f(s,t) = 0.12
\sim 0.087$ at $E^L_\gamma = 2 \sim 3$ GeV.
The idea of this paper is to extract the value of ${\cal R}$ from the
measurement of the polarization observables.
So in our qualitative estimation we take ${\cal R} = 0.1$ and we write
the real part of the VDM amplitude as
\begin{eqnarray}
\mbox{Re}\, T^{\rm VDM}_{f,i}= {\cal R}\, \mbox{Im}\, T^{\rm VDM}_{f,i}.
\end{eqnarray}
The VDM amplitude will be renormalized by multiplying
$1/\sqrt{1+{\cal R}^2}$.
Since ${\cal R}$ is around 0.1, the contribution from the real
part of VDM amplitude to the differential cross section is only about 1\%
of that of the imaginary VDM amplitude, which makes it hard to disentangle
the real part of VDM amplitude from the cross section measurements.
It should be kept in mind that this relation is valid only at
$|t| \to |t|_{\rm min}$ (or $\theta\to 0$) that is the most interesting
region where the cross section is at maximum.
However, for our qualitative analysis we will assume this relation for
the whole region of $t$ \cite{Deus}.
We also assume the constant value of ${\cal R}$ at low energy although
it is a function of $s$ in general.

It is, then, straightforward to obtain the corresponding amplitudes in
helicity basis with the relation \cite{JW,POL},
\begin{eqnarray}
  H_{\lambda_f,\lambda_\omega;\lambda_i,\lambda_\gamma} =
  (-1)^{1-\lambda_i-\lambda_f} \sum_{m_i,m_f}
  d^{1/2}_{m_i,-\lambda_i} (0)\,
  d^{1/2}_{m_f,-\lambda_f} (\theta)\,
  T_{m_f,\lambda_\omega;m_i,\lambda_\gamma},
\end{eqnarray}
where $\lambda_{i,f}$ are the helicity of the target and recoil proton,
respectively.
Analyses of the amplitudes show that at small $|t|$ their dominant parts
have the spin/helicity conserving form as
\begin{eqnarray}
\mbox{Im}\, H^{\rm VDM}_{\lambda_\omega,\lambda_f;\lambda_\gamma,\lambda_i}
&=& M^{\rm VDM}_0\,
\delta_{\lambda_\omega,\lambda_\gamma}\,
\delta_{\lambda_i,\lambda_f},
\label{VDM-A}\\
H^{\rm OPE}_{\lambda_\omega,\lambda_f;\lambda_\gamma,\lambda_i} &=&
2\lambda_i\lambda_\gamma M^{\rm OPE}_0
\delta_{\lambda_\omega,\lambda_\gamma}\,
\delta_{\lambda_i,\lambda_f},
\label{OPE-A}
\end{eqnarray}
where
\begin{eqnarray}
M^{\rm VDM}_0 &\simeq& -2|\vec{k}| C T_{0}(1+ \alpha\alpha'),  \nonumber \\
M^{\rm OPE}_0 &\simeq& -\frac{\nu (E_\omega-|\vec{q}|)}
{t - M_\pi^2} g_{NN\pi} \tilde g_{\omega\pi\gamma}.
\end{eqnarray}
The qualitative difference between (\ref{VDM-A}) and (\ref{OPE-A}) lies
on the existence of the additional phase factor $2\lambda_i\lambda_\gamma$
in (\ref{OPE-A}) that comes from the $NN\pi$ coupling and the magnetic
structure of the $\omega\pi\gamma$ interaction.
This factor plays an important role in the longitudinal double polarization
asymmetry $L_{\rm BT}$ for the circularly polarized photon beam defined as
\begin{eqnarray}
L_{\rm BT}=
\frac{{d\sigma}_\rightrightarrows - {d\sigma}_\rightleftarrows}
     {{d\sigma}_\rightrightarrows + {d\sigma}_\rightleftarrows},
\end{eqnarray}
where ${d\sigma}_\rightrightarrows$ (${d\sigma}_\rightleftarrows$)
represents $d\sigma/dt$ for the parallel (anti-parallel) helicity states
of the target and the photon beam.
Near the forward scattering region, therefore, we have
\begin{eqnarray}
L_{\rm BT} &\simeq&
\frac
{\Bigl\vert iM^{\rm VDM}_0+ {\cal R} M^{\rm VDM}_0-M^{\rm OPE}_0\Bigl\vert^2
-\Bigl\vert iM^{\rm VDM}_0+ {\cal R} M^{\rm VDM}_0+M^{\rm OPE}_0\Bigl\vert^2}
{\Bigl\vert iM^{\rm VDM}_0+ {\cal R} M^{\rm VDM}_0-M^{\rm OPE}_0\Bigl\vert^2
+\Bigl\vert iM^{\rm VDM}_0+ {\cal R} M^{\rm VDM}_0+M^{\rm OPE}_0\Bigl\vert^2}
\nonumber\\
&\simeq& -2\, {\cal R} \, \eta_\omega
\sqrt{\frac{d\sigma^{\rm OPE}\, d\sigma^{\rm VDM}}{d\sigma^{\rm tot}}},
\label{LBT0}
\end{eqnarray}
for small value of ${\cal R}$, where $\eta_\omega$ is the sign of
$\tilde g_{\omega\pi\gamma}$ and $d\sigma^{\rm OPE}$ denotes the
differential cross section $d\sigma/dt$ of OPE, etc.
This shows that $L_{\rm BT} (\theta=0)$ vanishes when ${\cal R}=0$ and
its sign is directly related to the phase $\eta_\omega$.

\begin{figure}[t]
\centering
\epsfig{file=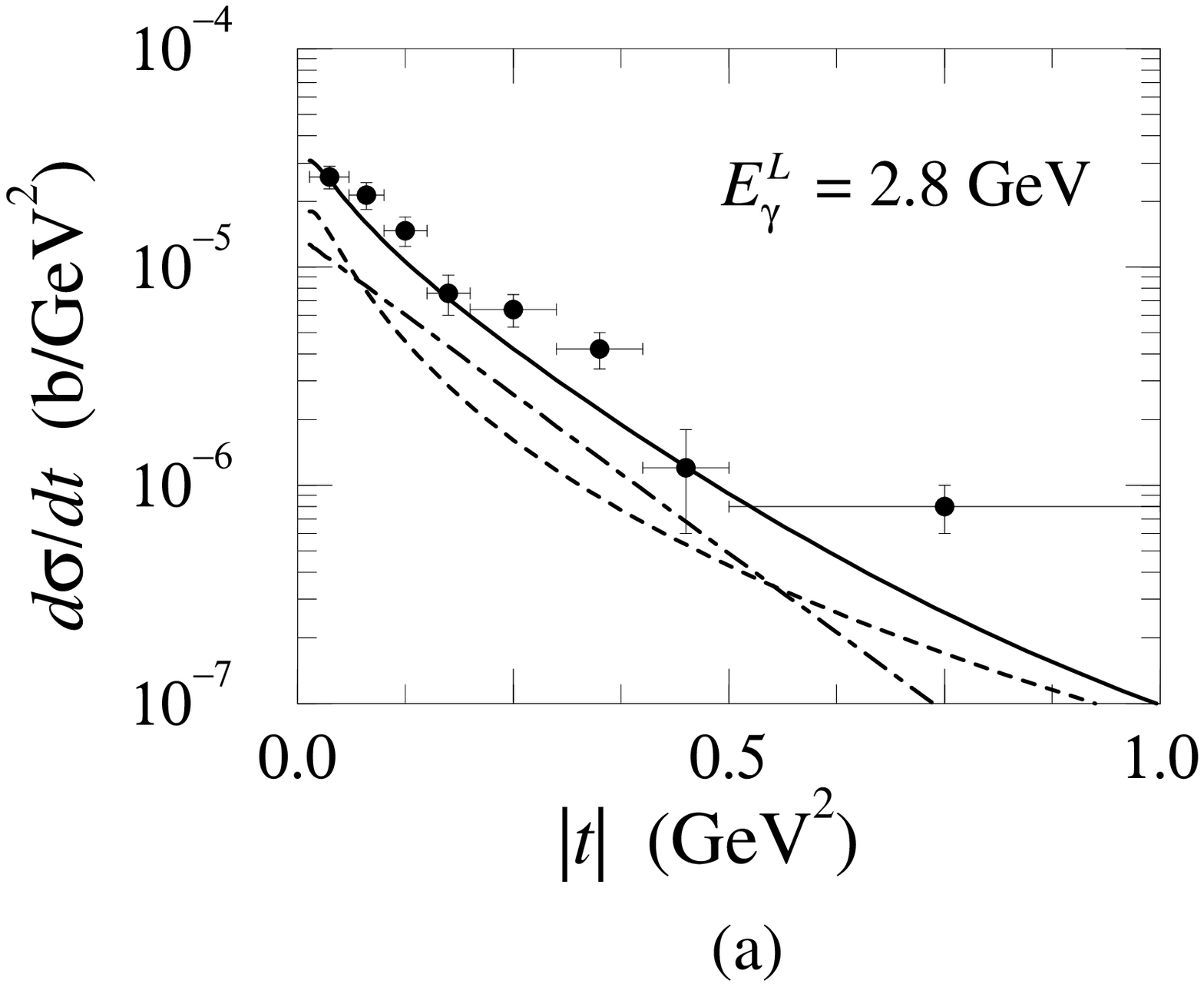, width=7cm} \quad
\epsfig{file=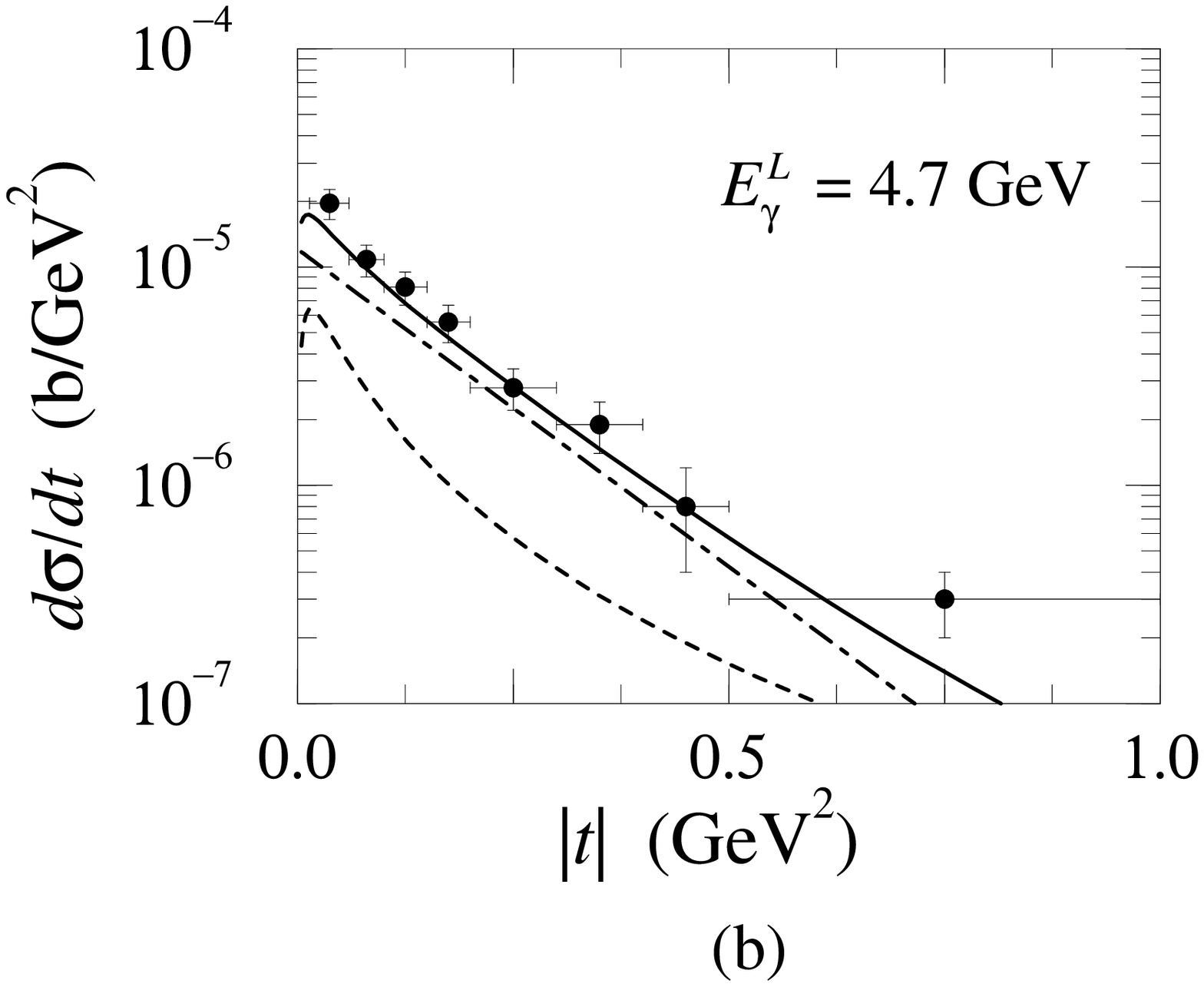, width=7cm}
\caption{Unpolarized $\omega$ photoproduction cross sections at two
         photon energies, (a) $E_\gamma^L=2.8$ GeV and (b) 4.7 GeV.
         The dotted and dot-dashed lines correspond to the OPE and VDM
         contributions, respectively, and the solid lines are the total
         differential cross section.
         The experimental data are taken from \protect\cite{Ballam73}.}
\label{fig:UN}
\end{figure}

\begin{figure}
\centering
\epsfig{file=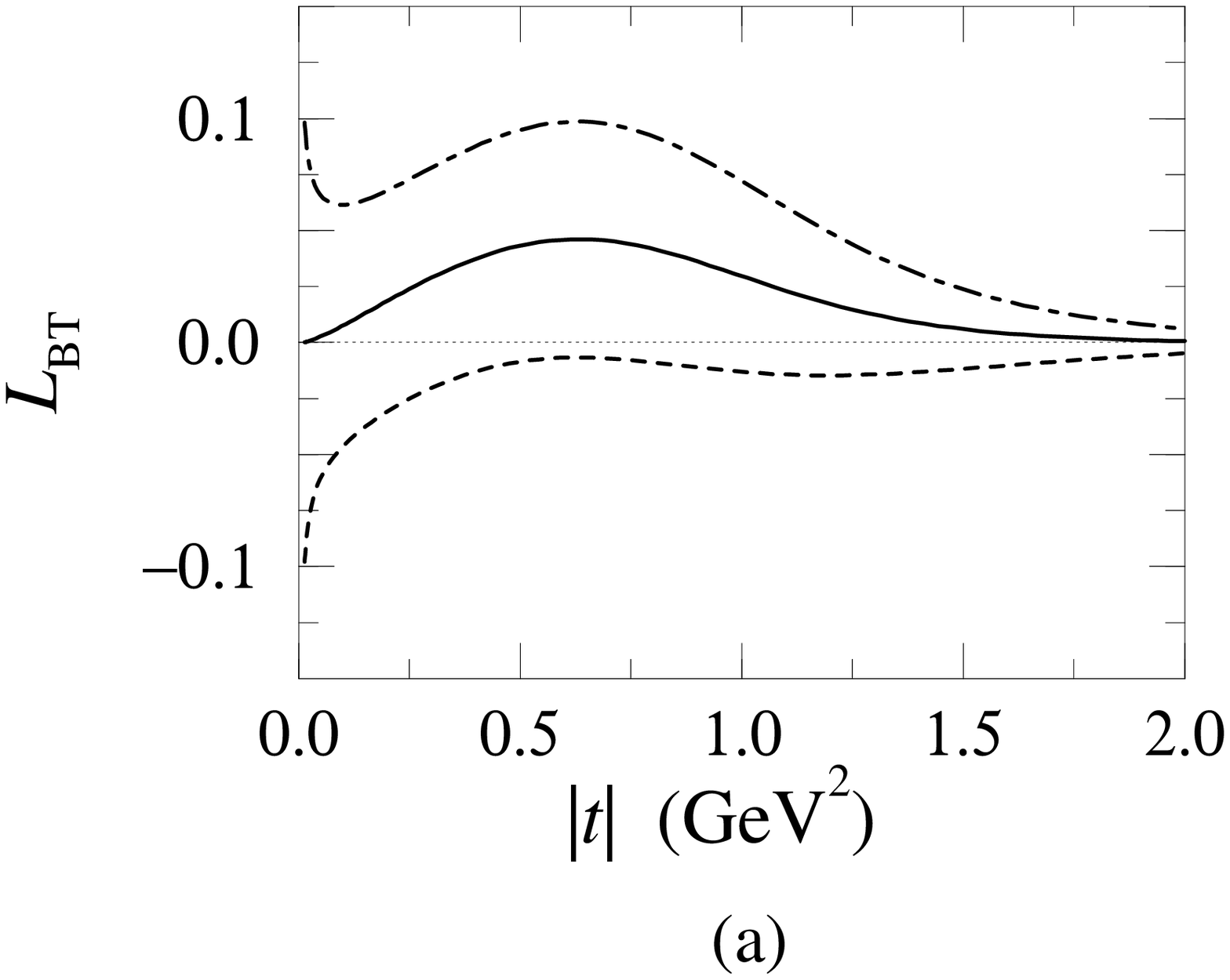, width=7cm} \quad
\epsfig{file=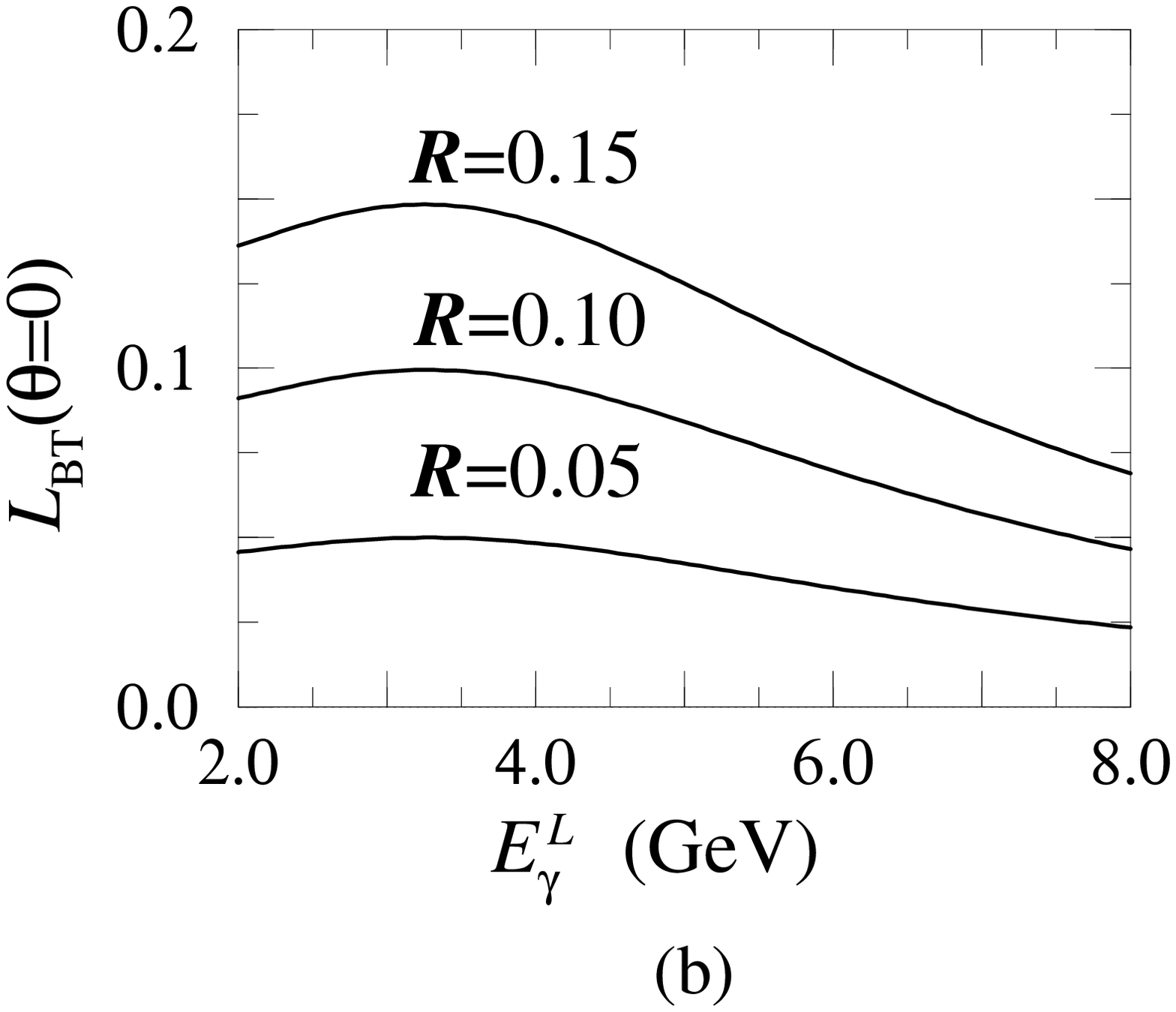, width=7cm}
\caption{Double polarization asymmetry $L_{\rm BT}$:
    (a) $t$-dependence at fixed energy $E_\gamma^L = 2.8$ GeV.
    The solid line is the result {\it without} $\mbox{Re}\,T^{\rm VDM}$
    and the dotted (dot-dashed) line is for $\eta_\omega=+1$ ($-1$)
    {\it with} $\mbox{Re}\,T^{\rm VDM}$ when ${\cal R}=0.1$.
    (b) $E_\gamma^L$-dependence at $|t|=|t|_{\rm min}$
    (i.e., $\theta=0$) with various ${\cal R}$ for $\eta_\omega=-1$.
    The results for $\eta_\omega=+1$ can be obtained by changing the sign
    of $L_{\rm BT}(\theta=0)$.
    The OPE amplitude is included in all graphs.}
\label{fig:BT}
\end{figure}

Results of our calculation are shown in Figs. \ref{fig:UN} and
\ref{fig:BT}.
Shown in Fig. \ref{fig:UN} are the unpolarized cross sections at two
initial photon energies, 2.8 and 4.7 GeV, with the experimental data
from Ref. \cite{Ballam73}.
One can find that the VDM and OPE channels have the same order of
magnitude at smaller $E_\gamma^L$ and similar $t$-dependence so that
they cannot be distinguished easily from the comparison with the data.
Furthermore, the dependence of the cross section on the phase
$\eta_\omega$ is negligible because of small value of ${\cal R}$.
Predictions for the double polarization asymmetry $L_{\rm BT}$ in the
model of VDM and OPE are shown in Fig. \ref{fig:BT}.
Its $t$-dependence is given in Fig. \ref{fig:BT}(a) with the initial
photon energy $E_\gamma^L = 2.8$ GeV, where the solid line shows the
prediction with the pure imaginary VDM amplitude (and with OPE) and
the dotted (dot-dashed) line corresponds to $\eta_\omega=+1$ ($-1$)
including the real part of the VDM amplitude with ${\cal R}=0.1$.
It shows the non-monotonic behavior and some enhancement at small $|t|$.
In Fig. \ref{fig:BT}(b), we give the $E_\gamma^L$-dependence of
$L_{\rm BT}$ for several values of ${\cal R}$ at $|t|_{\rm min}$.
This shows that the magnitude of the asymmetry $L_{\rm BT}$ is proportional
to the real part of the VDM amplitude so that it vanishes with ${\cal R}=0$
at $\theta=0$.
It also shows that the asymmetry decreases with increasing photon energy
and its sign at the forward scattering region is directly related to the
phase $\eta_\omega$.

In summary, we find that the double polarization asymmetry $L_{\rm BT}$,
especially at forward scattering region, is very sensitive to the real
part of the VDM diffractive photoproduction amplitude and the phase of
the $\omega\pi\gamma$ coupling constant.
This indicates that the both (${\cal R}$ and $\eta_\omega$) can be
directly extracted from the double polarization measurements.
Since the asymmetry decreases with increasing photon energy, the optimal
initial photon laboratory energy would be less than 5 GeV.
The currently available experimental data \cite{Ballam73} are not sufficient
for their estimates and new experiments are strongly favored in the new
facilities at the photon energy region of $2 \sim 4$ GeV
(see, e.g., Ref. \cite{Fujiwara}).
At theoretical side, further refinement of the photoproduction mechanisms
is also required.
For example, we need more precise information of the VDM amplitude to
determine the sign and magnitude of ${\cal R}$ in a more realistic way,
because it depends on the energies and the parameterization of the
imaginary VDM amplitude \cite{SS75}.
It would be also interesting to study the final state interactions, which
can contribute to the polarization observables at this energy region.
Finally, the phase fixing of OPE amplitude in this manner can give a clue
to the phase problem of pion photoproduction \cite{GG93,FM97,ITC}.

We acknowledge the fruitful discussions with M. Fujiwara, T. Kinashi,
H. Toki and S.N. Yang.
One of us (Y.O.) is grateful to the Alexander von Humboldt Foundation
for financial support.

\rule{\textwidth}{0.7mm}

\end{document}